\documentclass[letterpaper,twocolumn,10pt]{article}

\usepackage{usenix-2020-09}

\usepackage{tikz}
\usepackage{amsmath}
\usepackage{pifont}  % For checkmark and cross symbols
\usepackage{array}   % For customizing column alignment
\usepackage{float}  % For positioning of images, tables
\usepackage{footnote}
\usepackage{authblk}
\usepackage{courier} %% Sets font for listing as Courier.
\usepackage{listings}
\usepackage{xcolor}
\newcommand{\explorerr}{Cryptoscope} % Can change to QS Explorer
\newcommand{\explorer}{Cryptoscope } % Can change to QS Explorer

\begin{document}
%-------------------------------------------------------------------------------

%don't want date printed
\date{}

\title{\Large \bf \explorerr: Analyzing cryptographic usages in modern software}
% Thanks again to chatGPT

%for single author (just remove % characters)

% %for single author (just remove % characters)
% \author{
% {\rm Your N.\ Here}\\
% Your Institution
% \and
% {\rm Second Name}\\
% Second Institution
% % copy the following lines to add more authors
% % \and
% % {\rm Name}\\
% %Name Institution
% } % end author

% using authblk
\author{Micha Moffie}
\author{Omer Boehm}
\author{Anatoly Koyfman}
\author{Eyal Bin}
\author{Efrayim Sztokman}
\author{Sukanta Bhattacharjee}
\author{Meghnath Saha}
\author{James McGugan}
\affil{IBM Research} 
% \author[1]{James McGugan}
% \affil[1]{IBM Research}

% get full details:
% + // Anatoly Koyfman, anatoly@il.ibm.com
% + // Eyal Bin, bin@il.ibm.com
% + // "Sukanta Bhattacharjee" <Sukanta.Bhattacharjee@ibm.com>
% + // "Meghnath Saha" <meghsaha@in.ibm.com>
% + // Omer Boehm, omerb@il.ibm.com
% // James McGugan, James.McGugan@ibm.com
% + // Efrayim Sztokman, efrayim.sztokman@ibm.com
% + // Micha

% < 1
% Initial paper submissions ...  should consist of at most 13 typeset pages for the main body of the paper, one additional page for discussing ethics considerations and compliance with the open science policy, and a bibliography and well-marked appendices. At submission time, there is no limit on the length of the bibliography and appendices but reviewers are not required to read any appendices. 
% https://www.usenix.org/conference/usenixsecurity25/submission-policies-and-instructions

\maketitle

\begin{abstract}

The advent of quantum computing poses a significant challenge as it has the potential to break certain cryptographic algorithms, necessitating a proactive approach to identify and modernize cryptographic code. Identifying these cryptographic elements in existing code is only the first step. It is crucial not only to identify quantum vulnerable algorithms but also to detect vulnerabilities and incorrect crypto usages, to prioritize, report, monitor as well as remediate and modernize code bases. A U.S. government memorandum require agencies to begin their transition to PQC (Post Quantum Cryptograpy) by conducting a prioritized inventory of cryptographic systems including software and hardware systems.
%For many organizations, this means not only complying with regulatory requirements but also mitigating risks associated with outdated or vulnerable cryptographic implementations, ultimately ensuring the resilience and security of their operations in an evolving technological landscape.

In this paper we describe our code scanning tool - \emph{\explorer}- which leverages cryptographic domain knowledge as well as compiler techniques to statically parse and analyze source code. By analyzing control and data flow the tool is able to build an extendable and querriable inventory of cryptography. \explorer goes beyond identifying disconnected cryptographic API's and instead provides the user with an inventory of \emph{cryptographic assets} - containing comprehensive views of the cryptographic operations implemented. We show that for more than 92\% of our test cases, these views include the cryptographic operation itself, APIs, as well as the related material such as keys, nonces, random sources etc. Lastly, building on top of this inventory, our tool is able to detect and report \emph{all} the cryptographic related weaknesses and vulnerabilities (11 out of 15) in CamBench - achieving state-of-the-art performance.

% Restore keywords when publishing
% keywords: Quantum safe cryptography, cryptographic inventory, CBOM, 
% cryptographic vulnerabilities, cryptographic misuse, 
% cryptographic agility.
\end{abstract}

\section{Introduction}

\begin{figure}[!ht]
    \centering
        \includegraphics[width=0.5\textwidth]{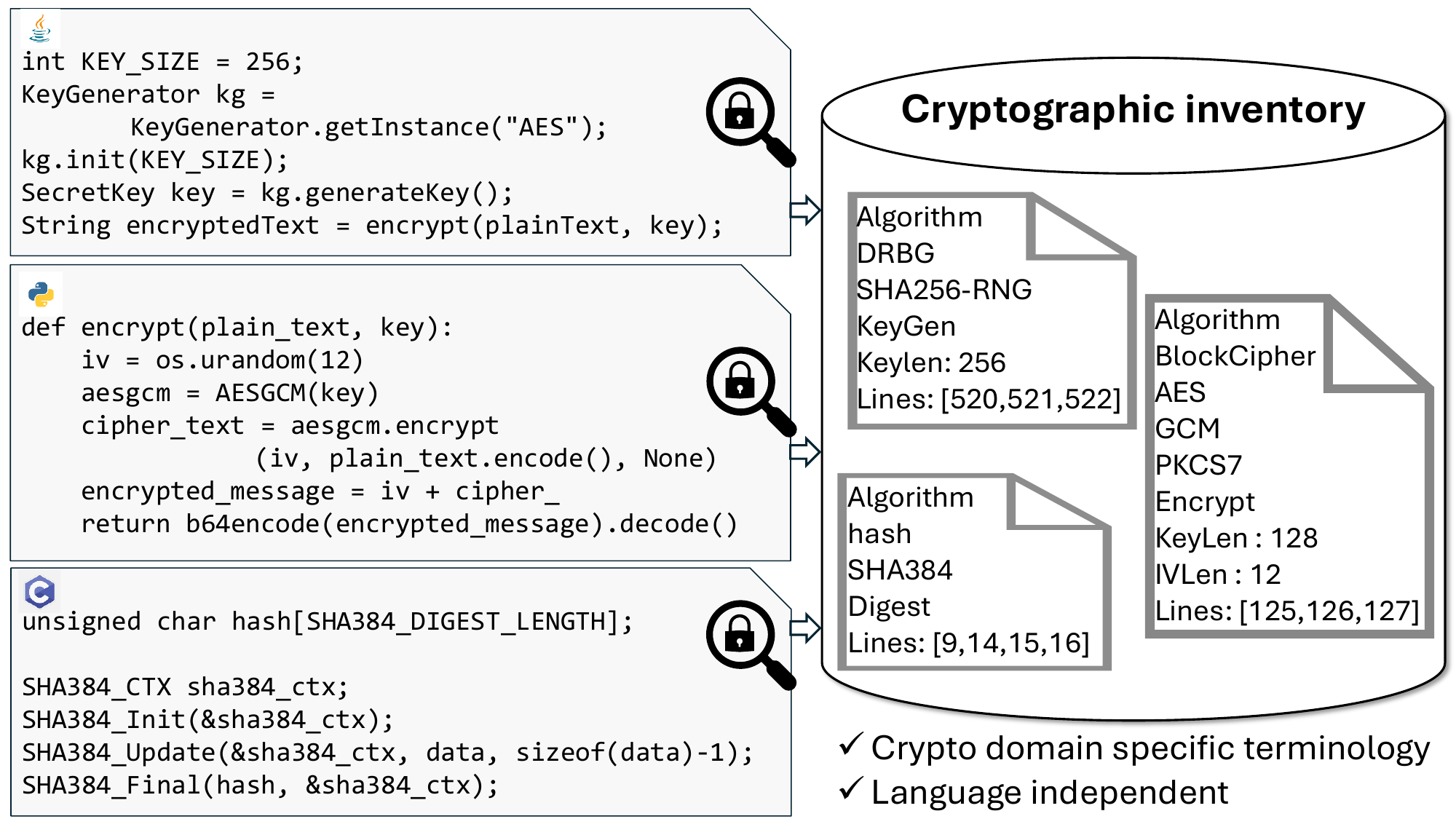}            
    \caption{\label{fig:problem_statement} 
    Illustrating the expected input and output to clarify challenge: first, identify the implementation of cryptographic operations in source code across different programming languages in a generic manner, second, represent the complete operational semantics in a unified way.
}
\end{figure}

The US National Security Memorandum~\cite{US-Memorandum} identifies the steps needed to address the risks posed by quantum computers and provides specific actions to be taken. Specifically, the memorandum requires inventoring cryptographic assets in deployed systems with the goal of identifying Quantum vulnerabilities and prioritizing the process of migrating the systems to quantum-resistant cryptography. Addressing such a requirement, in organizations with large code bases is not an easy task. Not only are applications written using a plethora of programming languages and API's, importantly, the resulting inventory of cryptography must be unified, complete and consistent across all applications regardless of the implementation details. Figure~\ref{fig:problem_statement} presents the expected input and output demonstrating the challenge. 

The requirements layed out in the memorandum are not without merit, the arrival of large-scale quantum computing offers great promise to science and society, but brings with it a significant threat to our global information infrastructure. Public-key cryptography - widely used on the internet today - relies upon mathematical problems that are believed to be difficult to solve given the computational power available now.
%and in the medium term.
However, popular cryptographic schemes, 
used in public key encryption, digital signatures, key establishments, etc., 
that are based on these hard problems – including RSA and Elliptic Curve Cryptography – will be easily broken by a quantum computer. This will rapidly accelerate the obsolescence of our currently deployed security systems and will have dramatic impacts on any industry where information needs to be kept secure.

% The NSA announced the Commercial National Security Algorithm (CNSA) Suite 2.0~\cite{CNSA2} to address the need for protection against a future deployment of a cryptanalytically relevant quantum computer (CRQC). CNSA 2.0 is a suite of Quantum Resistant (QR) algorithms approved for eventual National Security Systems (NSS) use. The suite lists the algorithms and their functions - including digital signing and key establishment - specifications, and parameters and provides a timeframe for its adoption by the end of 2033. 

The construction of an inventory of cryptographic operations (crypto assets) is essential to provide an organization with necessary information to support the process of migrating and modernizing cryptography to the quantum era. In particular, an inventory as shown in figure~\ref{fig:inventory_use_cases}, would be able to support:

%\begin{figure*}[!ht]
\begin{figure}[!ht]
    \centering
        \includegraphics[width=0.5\textwidth]{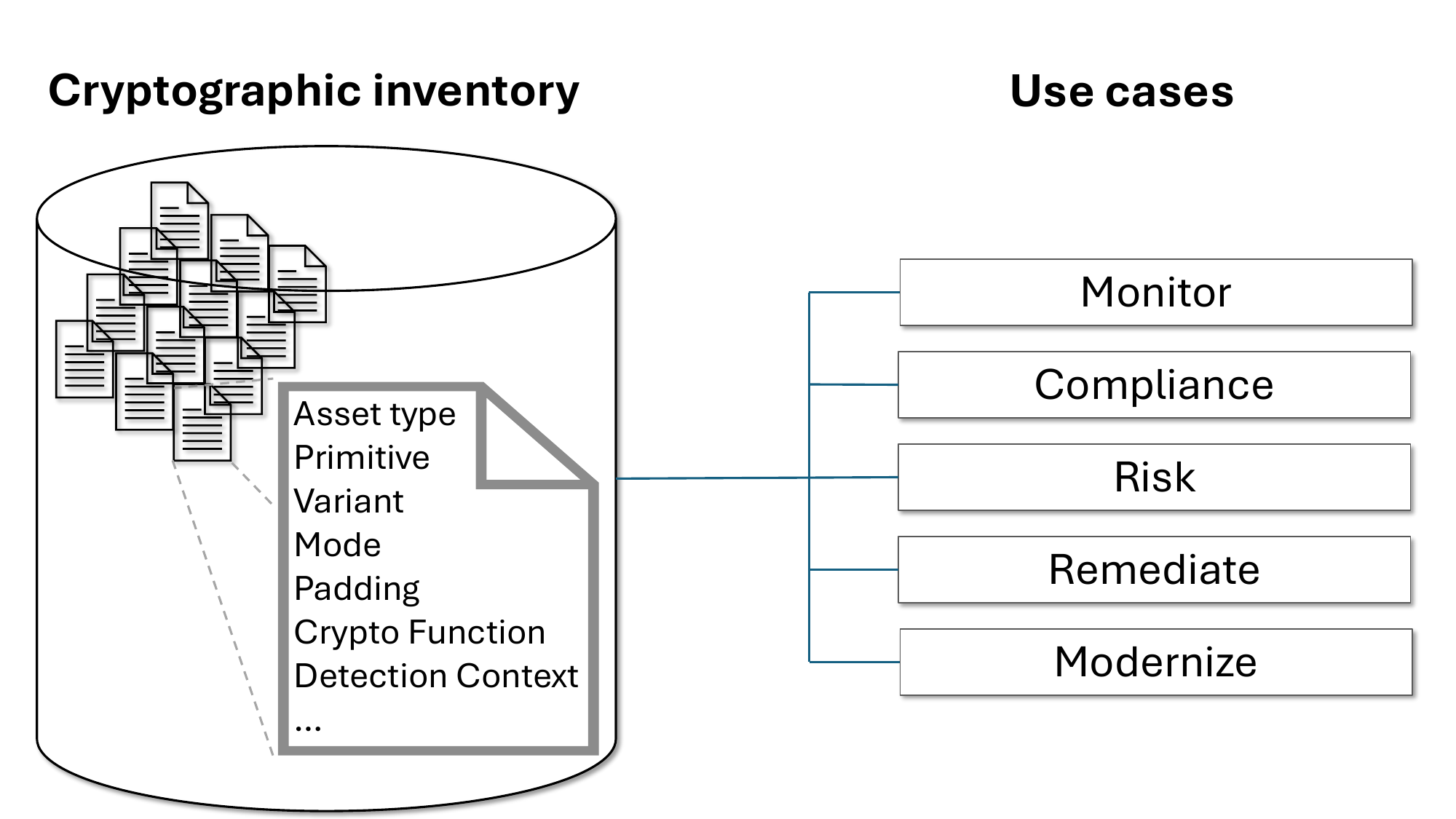}
    \caption{\label{fig:inventory_use_cases} The use cases that are supported by a complete and unified organizational cryptographic inventory.
}
\end{figure}
%\end{figure*}

\begin{itemize}
    \item Identification of existing cryptography across applications and continuous monitoring of cryptography usage within an organization
    \item Strengthen the ability to comply to different regulations and standards
    and support the enactment and enforcement of organizational wide cryptographic policies 
    \item Assess and prioritize the different applications based on risk and locate risky application that should be mitigated  
    \item Inform the process of application remediation with accurate findings (e.g. cryptographic use case, lines in code)  
    \item Provide organizational wide view to support modernization and insights to enable cryptographic agility across the organization
\end{itemize}
The effectiveness of such an inventory depends mainly on the \textbf{quality} of cryptographic assets listed. An accurate, complete and concise expression of the identified cryptography must account for all the different aspects of a cryptographic operation. Consider data encryption; in this case the algorithm, operation, mode, as well as related information such as the IV, %salt, 
key and source of randomness are required. In essence, a complete crypto asset will convey to the user the cryptographic operation \emph{semantics}. In addition, providing the evidences for these findings - e.g. lines of code - is essential both for validating the results and for supporting next steps such as remediation.
Importantly, incomplete or inaccurate representations of cryptographic operations semantics can result in a skewed view of the existing cryptography in the application as well as the organization as a whole, which could lead to inaccurate risk assessment, wrong prioritization and increased risk of continuously deployed weak cryptography.

In our work, we decouple the process of discovery (creating the inventory) from any specific use case. By providing as comprehensive a view as possible of the cryptographic usage in the application we expect to support a broad range of use cases.
Such a separation allows for independent implementation of any one of those use cases regardless of the discovery process specific realization.
%s, e.g. static vs. dynamic analysis. 
Specifically, we model rules for vulnerability identification using generic cryptographic terms (based on the inventory) rather than specific language and library APIs. Thus enabling a single implementation of vulnerability identification to support a wide range of programming languages.

Our main contributions in this work are as follows: (1) A generic, language agnostic, mechanism to discover crypto-assets in source code, (2) providing the user with complete, unified cryptographic operational semantics. And, 
% (in a form of a CBOM, Cryptographic Bill of Material). 
% there is a reference later.
(3) a robust method to extract crypto related vulnerabilities and missuses on top of the crypto-inventory. 
We evaluate our approach on real world applications and crypto-related benchmark and show it is able to correctly discover crypto assets and accurately identify all crypto-related vulnerabilities and missuses - showcasing the cryptographic operation semantic completeness contained in the crypto-assets.

% The rest of the paper is organized as follows; in the next section %section~\ref{sec:relatedArt} 
% we provide related art and background. %  on existing related tools and mechanisms. 
% Section~\ref{sec:Implementation} provides more details on our framework and how a crypto-assets are identified and represented. Evaluation on real world application is present in 
% Section~\ref{sec:evaluation}. In section~\ref{sec:vulneravilityDiscovery} we show how we can build on the inventory to identify crypto vulnerabilities. A short summary and next steps are presented in section~\ref{sec:summary}. 

\section{Overview}
\label{sec:overview}

%\subsection{Design motivation}

Implementing cryptographic algorithms and protocols often requires calling a few library API calls, usually in a predefined order, and provide specific parameters to those calls.
These parameters, such as algorithm name, are often hard-coded in the code base. 
%For example, parameters can be encoded in configuration files, as global values or even hard-coded as strings in call sites. 
Often, programmers will wrap cryptographic API's in utility functions and expose a more convenient "internal" API to other modules in the code. In many of those cases, the "internal" API is parameterized to allow for more control by the calling modules. For example, the key or algorithm name could be provided by a module calling the 'internal' API.
In such cases, it is essential to trace back, starting from the cryptographic API calls to the source of the cryptographic parameters (e.g. the call site to the 'internal' API) to be able to detect the exact cryptographic operation - and create a meaningful crypto-asset. Moreover, an additional complexity arises in cases where multiple calls to the 'internal' API exist - requiring us to trace the parameters back to multiple sources and possibly create multiple different crypto assets (differing only by the key size for example).

%which may depend on non-local inputs such as configuration files or configuration databases. Moreover, arguments for calls to cryptographic functions sometimes depend on other parts of the code which are in completely different parts of the application.

A straightforward method for identifying usage of cryptography is by simply identifying calls to standard cryptographic API's. This process is analogous to "grep" and results in low accuracy. For example, this method may identify the instantiation of an algorithm but would not be able to identify the key size it used or whether this algorithm was in fact used to perform decryption or encryption. More complex patterns could be introduced, however they will be limited to a predefined set of use cases and may not find all related API calls as only those specified within the pattern can be accounted for. Moreover, patterns, in addition to their modeling, also require maintenance and are likely to be programming language and cryptographic library API specific. These methods do not provide a generic, solid process to identify related API calls.

We are now ready to summarize the main requirements driving our design:
\begin{itemize}
    \item Trace crypto-parameters back to detect hard coded values such as "AES/CBC/PKCS5Padding" or key size. Also, support tracing back to multiple different (e.g. call) sites, as these may represent different crypto assets
    \item Provide a method to identify related cryptographic API's in a generic manner
\end{itemize}

Reviewing those requirements, we find static analysis can provide much support. Specifically:
\begin{itemize}
    \item Constant value analysis to identify hard coded values. For example, the size of the required key passed to the \texttt{KeyGenerator.init(..)} function).
    \item Program slicing allows  us to isolate, based on a critiria, part of the program statements while preserving a subset of the program behaviour. Intuitively, it allows us to identify, for example, a set of statements that would result in a particular assignment of value to a crypto parameter - while ignoring all other statements. In turn, the crypto asset derived from this specific value, will relate to this particular slice. In the example above, one could imagine each call to the "internal" API as a different slice of the program - exhibiting a particular behaviour summarized in a separate crypto asset. A thorough explanation on program slicing can be found in~\cite{ProgramSlicing}.
    \item Identify related API calls by tracing cryptographic objects (on which methods are invoked) as well as cryptographic parameters (which may be a result of other cryptographic calls) by using data and control flow analysis.
\end{itemize}
%Static analysis is typically used to check for functional bugs, vulnerabilities, or non-compliant coding, however, it can also be used to document and record cryptographic evidences.

In our design we make use of these exact static analysis algorithms to create program slices, identify related crypto API calls and find values of hard coded crypto parameters. We will provide more details in following sections.

Before we continue, we provide some background on the crypto assets defined in the CBOM (Cryptographic Bill of Material) - the building blocks of the inventory - and discuss several important issues.

\subsubsection*{Crypto Asset}

The Cryptography Bill of Materials (CBOM) is part of CycloneDX v1.6 standard~\cite{CBOM}.
Its goal is to extend the Software Bill of Materials (SBOM) - which was originally designed to be used in application security and supply chain component analysis - with the ability to express cryptography. 
%Specifically, the CBOM is an object model able to describe cryptographic assets and their dependencies. Effectively, the CBOM allows us to describe the cryptography in a language-independent manner.
The CBOM includes a new component of type \emph{crypto-asset} which includes a description of a usage of a cryptographic certificate, protocol or algorithm using terminology from the cryptography domain such as primitive, variant, mode, function etc. Specifically, a crypto asset will include:
\begin{itemize}
    \item All relevant crypto properties as well as related crypto material (such as private/public/secret keys, initialization vector, salt, digest, signature and password) - if any.
    \item The relevant context and evidences for the cryptographic operation - API calls, relevant crypto parameters and their locations in the code.
\end{itemize}
Our modeling is consistent with the CBOM standard definition. Currently, \explorer supports assets of type algorithm and related crypto material; protocols and certificates are left for a future release. 

\subsubsection*{Semantic completeness}
While the CBOM defines a way to describe, %convey, 
and store cryptographic information it does not require nor define the expected completeness or semantics of a crypto asset. In fact, the user of the CBOM has the flexibility to use the CBOM to describe solely a primitive (algorithm) such as a signature, based, for example, on a single cryptographic API. Or, alternatively, describe a more semantically complete operation containing a primitive (e.g. encryption), variant (e.g. AES), mode (GCM), operation (decrypt), key size (256) and the initialization vector - by analyzing multiple related cryptographic API's. Although both cases can result in a valid and legal CBOM, its contents and quality can make a huge difference to downstream users; the more complete the description is - the better policy enforcement, vulnerability identification, and support for remediation would become.

%In fact, the discovery process addresses the usage of crypto algorithms and the related crypto material. There are different types of crypto asset properties (\emph{Crypto properties}) addressing different aspects of cryptography including variant, primitive, mode etc. Each one of those crypto properties types contain the relevant value (if available) and context. For example, \explorer supports the crypto property of type function with the following values: \emph{Encrypt}, \emph{Decrypt}, \emph{Digest}, \emph{Tag}, \emph{Sign}, \emph{Verify}, \emph{Encapsulate},\emph{Decapsulate}, \emph{Random number generation}, \emph{Key derivation}, and \emph{Key generation} as defined in CBOM.

%\emph{Related crypto material} is additional information that is related to a crypto asset. It contains a predefined type (private key, public key, seed, etc.), value (if available) and context. Unlike crypto property of an asset (e.g. private key size), crypto material provides the actual data element used in construction (e.g. the private key value). This is consistent with the CBOM standard definition. 

%\todo[inline]{related crypto material can be a nonce, key etc. merge the above this is like a CBOM. there is no significant difference between related and property...}

\subsubsection*{Language independence}

Crypto assets are effectively language-independent since all the crypto properties and related crypto materials use terminology taken from the domain of cryptography. This is an important property as it (1) provides for a unified view regardless of the input programming language and (2) abstracts away implementation details allowing for later consumers e.g. policy enforcement tools or vulnerability detectors etc. to focus on the cryptography. For example, insecure usage of SHA-1, a known vulnerable message digest algorithm, can be easily identified using a single check of the asset property "variant", rather than accounting for a particular programming language and specific library API implementations.

\subsubsection*{Meaningfull crypto assets}

To illustrate and clarify what a crypto asset is, we provide examples - code snippets - and discuss the expected results. As stated above, our goal is to discover cryptographic operations. Therefore, our guiding principal is to identify that a cryptographic action has actually been performed. Specifically, to obtain a meaningful expression of the usage of cryptography, we differentiate between "Usage of a cryptographic algorithm” and "Usage of cryptographic library API calls". For example, 
\texttt{cipher.getInstance(”AES”)} \textbf{will not} - by itself - be considered a crypto asset unless \texttt{cipher.doFinal()} is followed later in the code - resulting in encryption (or decryption) to be performed.

Moreover, since software is built in abstraction layers - each layer may abstract some of the crypto details - we identify the asset in one single layer. 
In particular, the cryptographic function calls identified as part of an asset will be those belonging to a standard library such as JCA~\cite{jca_doc} or BouncyCastle~\cite{bouncycastle}. (These are also encoded in our Knowledge base.)
Barring such a definition, the discovery process could theoretically identify crypto assets for the same crypto operation multiple times across different abstraction layers.

Consider the two code segments in listing~\ref{listing:code_with_asset} - the first contains a "sign" operation while the second is performing "key generation" operation (using JCA API's). Both of these  \textbf{will} result in crypto assets:
%\begin{listing}[hb]
% \begin{minted}[
%     linenos=false,          % Show line numbers
%     frame=lines,           % Draw a frame
%     fontsize=\small,      
%     breaklines=true,       % Allow breaking long lines
% ]{java}
% Signature ecdsa = Signature.getInstance
%                         ("SHA256withECDSA");
% ecdsa.initSign(key); 
% ecdsa.update(cri.toByteArray()); 
% ecdsa.sign();
% \end{minted}
\begin{lstlisting}[language = Java, frame = tb, numbers=none]
Signature ecdsa = Signature.getInstance("SHA256withECDSA");
ecdsa.initSign(key); 
ecdsa.update(cri.toByteArray()); 
ecdsa.sign();
\end{lstlisting}

% \begin{minted}[
%     linenos=false,          % Show line numbers
%     frame=lines,           % Draw a frame 
%     fontsize=\small,      
%     breaklines=true,       % Allow breaking long lines
% ]{java}
% KeyPairGenerator keyPairGen = KeyPairGenerator.getInstance("RSA"); 
% keyPairGen.initialize(2048);
% KeyPair keyPair = keyPairGen.generateKeyPair(); 
% \end{minted}
\begin{lstlisting}[language = Java, frame = tb, numbers=none,
captionpos=b,
caption = {Code samples where JCA API's are called, a crypto operation is performed and a crypto asset will be created.},
label = {listing:code_with_asset}
]
KeyPairGenerator pairGen = KeyPairGenerator.getInstance("RSA");
pairGen.initialize(2048);
KeyPair keyPair = pairGen.generateKeyPair(); 
\end{lstlisting}
% \caption{Code samples where JCA API's are called, a crypto operation is performed and a crypto asset will be created}
% \label{listing:code_with_asset}
%\end{listing}

In contrast, listing~\ref{listing:code_without_asset_I} shows code snippets that \textbf{will not} contain crypto assets. The first snippet constructs a certificate from a certificate stream while the second snippet is reading an existing key from a key store. In both cases no cryptographic operations are involved as they have already been performed prior to the reading of the data.

%\begin{listing}[h]
% \begin{minted}[
%     linenos=false,          % Show line numbers
%     frame=lines,           % Draw a frame 
%     fontsize=\small,      
%     breaklines=true,       % Allow breaking long lines
% ]{java}
% CertificateFactory factory = CertificateFactory.getInstance("X.509");
% factory.generateCertificate(certificateStream);
% \end{minted}
\begin{lstlisting}[language = Java, frame = tb, numbers=none]
CertificateFactory factory = CertificateFactory.getInstance("X.509");
factory.generateCertificate(certificateStream);
\end{lstlisting}

% \begin{minted}[
%     linenos=false,          % Show line numbers
%     frame=lines,           % Draw a frame 
%     fontsize=\small,      
%     breaklines=true,       % Allow breaking long lines
% ]{java}
% KeyStore ks = KeyStore.getInstance("PKCS12"); 
% PrivateKey tmpPriv = pkEntry.getPrivateKey();
% \end{minted}
\begin{lstlisting}[language = Java, frame = tb, numbers=none,
captionpos=b,
caption = {Code samples where JCA API's are called - but no crypto is performed. No crypto asset will be created.},
label = {listing:code_without_asset_I}
]
KeyStore k = KeyStore.getInstance("PKCS12"); 
PrivateKey tmpPrv = pkEntry.getPrivateKey();
\end{lstlisting}
% \caption{Code samples where JCA API's are called - but no crypto is performed. No crypto asset will be created.}
% \label{listing:code_without_asset_I}
% \end{listing}

% The actual creation of a key and using it for a certificate would have happened prior to these examples and may have looked something like this:
% \begin{minted}[
%     frame=lines,           % Draw a frame 
%     fontsize=\small,      
%     breaklines=true,       % Allow breaking long lines
% ]{java}

% KeyPairGenerator keyPairGenerator = KeyPairGenerator
%     .getInstance("RSA");
% keyPairGenerator.initialize(2048);
% KeyPair keyPair = keyPairGenerator.generateKeyPair();

% ContentSigner signer = new JcaContentSignerBuilder("SHA256withRSA")
%     .build(keyPair.getPrivate());
% X509CertificateHolder certificateHolder = builder.build(signer);
% return new JcaX509CertificateConverter()
%     .getCertificate(certificateHolder);

% \end{minted}

Listing~\ref{listing:code_without_asset_II} shows another example where code abstracting standard API calls and therefore \textbf{will not} result in crypto assets. 
As mentioned above, only standard crypo library API calls are identified.(Similarly,
% Non standard code, e.g. code containing abstraction layers, 
custom, homegrown, cryptographic implementations will not be identified in the current implementation.) 
%But, any additional calls to \texttt{sha256} method of the \texttt{Hash} %class will not be registered as an Asset as we consider it already covered:
Note, that the implementation of the  
\texttt{MyHash.sha256} method may contain an asset. 

%\begin{listing}[h]
% \begin{minted}[
%     linenos=false,
%     frame=lines,           % Draw a frame 
%     fontsize=\small,      
%     breaklines=true,       % Allow breaking long lines
% ]{java}
% String text = "my bytes";
% MyHash hash = new MyHash();
% bytes digest = hash.sha256(text.getBytes());
% \end{minted}
\begin{lstlisting}[language = Java, frame = tb, numbers=none,
captionpos=b,
caption = {Code sample where JCA API's are abstracted. No crypto asset will be created for the code shown.},
label = {listing:code_without_asset_II}
]
String text = "my bytes";
MyHash hash = new MyHash();
bytes digest = hash.sha256(text.getBytes());
\end{lstlisting}
% \caption{Code sample where JCA API's are abstracted. No crypto asset will be created for the code shown.}
% \label{listing:code_without_asset_II}
% \end{listing}

%%% the complete code:
% import java.security.MessageDigest;
% ...
% public class MyHash {
%     public static final String HASH = "SHA-256";

%     public static byte[] sha256(byte[] input) {
%         // Instantiate and call MessageDigest digest()
%        ...
%     }
    
%     public static int main() {
%         String text = "my bytes";
%         MyHash hash = new MyHash();
%         bytes digest = hash.sha256(text.getBytes());
%    }

\section{Implementation}
\label{sec:Implementation}
% - Why not to start with "high-level overview" of your code scanning tool? The tool consists of X components: ... Component A scans the source code, B generates a graph, ...

%\subsection{\explorer design and implementation}

\begin{figure}[!ht]
    \centering
        \includegraphics[width=0.5\textwidth]{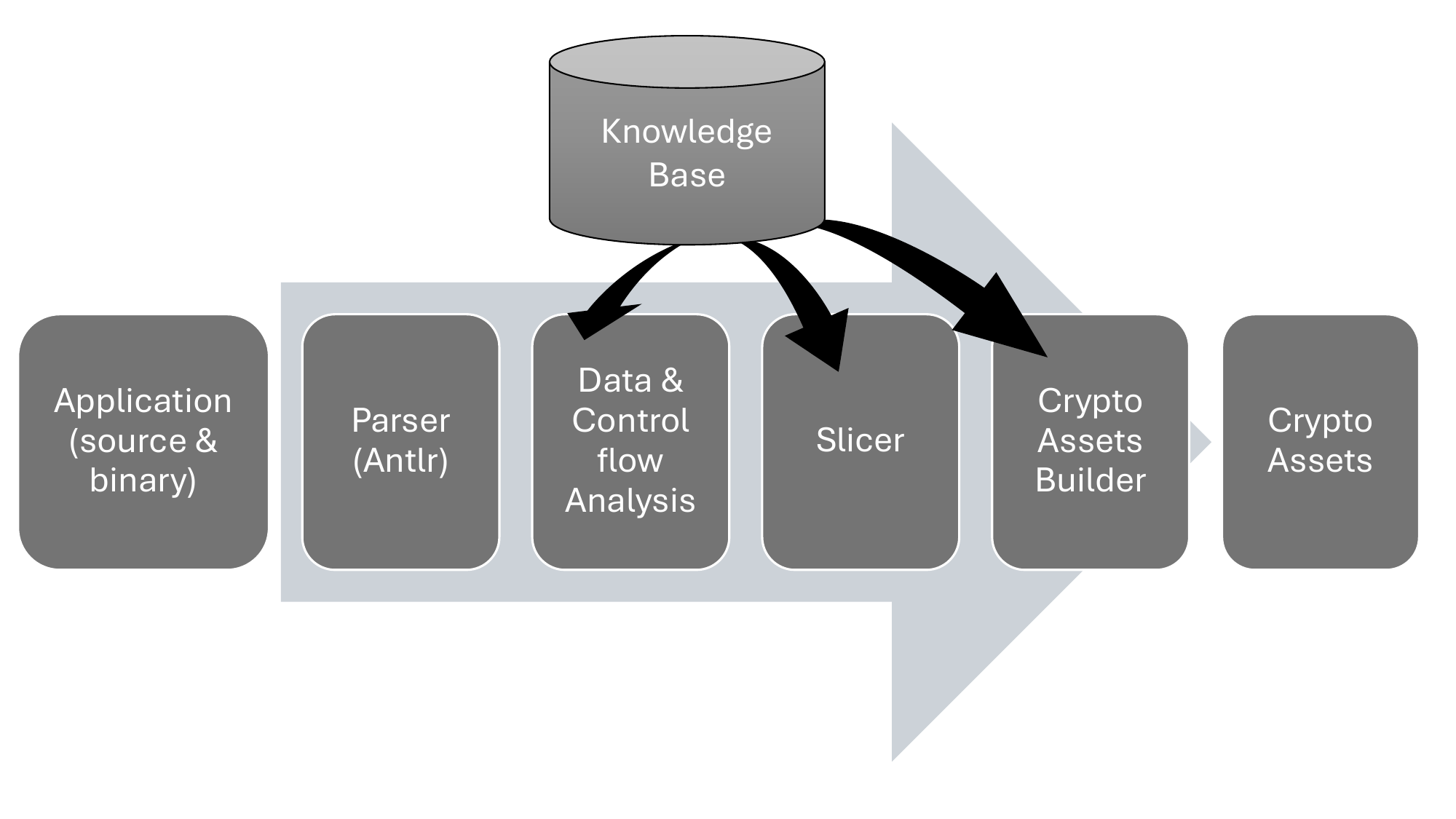}
    \caption{\label{fig:design} Crypto-asset discovery. The four stages of the static analysis pipeline: parsing the code, analyzing control and data flow, building slices based on crypto relevant criteria and finally constructing crypto assets for each slice.
            }
\end{figure}

A high level overview of \explorer is shown in figure~\ref{fig:design} and is composed out of 4 stages.
The discovery process begins with source code (and binaries) and ends with the discovered crypto-assets.

In the first stage, source code is tokenized and parsed (using Antlr~\cite{Antlr}) into an AST (Abstract Syntax Tree). The AST is used in the next stage to analyze the program blocks relationships and build in-memory data flow, control flow and call graphs. 
These constructs are further used in later stages.
Depending on the programming language, data types are identified in this stage (e.g. Java) or in the following stage (e.g. Python).

The Slicer performs backwards slicing on all statements in the code with matching slicing 
criterion\footnote{Note, the slicing is performed on high level code requiring some adjustments to the algorithms} (encoded in the knowledge base.). 
The slicing criteria is defined as an API call which completes the sequence of crypto calls and is the one that performs the operation such as \texttt{Cipher.doFinal()}, %\texttt{keyPairGen.generateKeyPair()} 
or \texttt{signature.verify(sig)}\footnote{The slicing criteria can be defined differently, in some cases it will require forward slicing or both backwards and forward slicing.}.
The Slicer identifies all program statements that influence the selected slicing criterion. The collected statements are divided according to their context, i.e. the control flow that led to them, and are enriched with inferred values. Each such "contextual" set of statements along with their corresponding arguments values are used to construct a slice.

The last stage constructs crypto assets. Here, each slice is analyzed separately. The slice 
already includes - by definition - all the relevant pieces of information for the cryptographic operation and once crypto properties are collected (using our knowledge base) - a complete crypto asset can be constructed.
Specifically, as mentioned above, the analysis is required to (1) identify values of the arguments of cryptographic API’s and (2) ‘string together’ related API calls. 
To address the first we employ interprocedural constant value analysis to derive constant values.  
To address the second, we first define related API calls as follows:
(I) A result of one API call is passed to (used by) another API call as a parameter.
(II) API methods are called using the same instance/object.
We then detect these two relationships on each slice by tracing data flowing through the program statements . 

The resulting crypto asset contains all the information that was required for the execution of the cryptographic algorithm, e.g., an encryption algorithm that uses a cipher instantiation method, cipher initialization method, secret keys generation methods, and the actual encryption operation method.
Similarly, a signature algorithm will include an instantiation using schema, the generated keys, and the actual sign or verify operations.

%`Static analysis identifies \textbf{every} possible realization of cryptography. The absence of the ability to tell which of those are actually going to be executed may result in false positives.

\subsubsection*{Knowledge base}

%The algorithm results in a set of crypto assets that can be represented in various output formats. 
%The knowledge base encapsulates all the required information about the various cryptographical libraries APIs, and provides, for each API call, a programming-language independent crypto properties set. This allows the tool to comply with the open-close principle, such that the KB hides all library-dependent details from the code, allowing it to work seamlessly on various implementations.      

The knowledge base (KB) contains all the information required to support the various steps during the discovery process. Importantly, all the required information is described using cryptographic terms that are independent of the programming language or the specific cryptographic library being used. 

At the heart of the KB lie the descriptions of cryptographic API calls %(crypto functions) 
from standard cryptograhic libraries in different programming languages. The description includes the function signatures and the identification of the crypto relevant parameters. In addition, the KB includes the \emph{association} between the cryptograpic calls and their parameter values and the cryptographic properties - effectively encoding the relevant cryptographic operational semantics.

Consider the simple code segment in listing~\ref{listing:code_get_instance}.
%\begin{listing}[h]
% \begin{minted}[
%     linenos=false,          % Show line numbers
%     frame=lines,           % Draw a frame 
%     fontsize=\small,      
%     breaklines=true,       % Allow breaking long lines
% ]{java}
% ...
% Cipher c = Cipher.getInstance("AES/GCM/NoPadding");
% ....
% \end{minted}
\begin{lstlisting}[language = Java, frame = tb, numbers=none,
captionpos=b,
caption = {A simple call to one of JCA's Cipher.getInstance API.},
label = {listing:code_get_instance}
]
...
Cipher c = Cipher.getInstance("AES/GCM/NoPadding");
...
\end{lstlisting}
% \caption{A simple call to one of JCA's Cipher.getInstance API.}
% \label{listing:code_get_instance}
% \end{listing}
The \texttt{getInstance()} method of Cipher class in JCA~\cite{jca_doc} has the following alternatives:
%There are several crypto functions named \texttt{getInstance}
\begin{itemize}
\item \texttt{getInstance(String transformation)}
\item \texttt{getInstance(String transformation, Provider provider)}
\item \texttt{getInstance(String transformation, String provider)}
\end{itemize}
Each one will be encoded separately in the KB. The first parameter in all the alternatives will be identified as crypto relevant and marked as a crypto parameter containing "transformation" semantics. Next, for each encoding of the transformation, e.g. "AES", "DES" or "AES/CBC/PKCS5Padding", we  associate the API name and the value of its first parameter with a list of predefined crypto properties. For example, \texttt{getInstance("AES")} will be associated with "blockcipher" (primitive property), "AES" algorithm (variant property),  "CBC" (mode property), "PKCS5" padding, as well as block size of 128 bits, and key size of 128 bits - all according to the documentation of the library.
The \texttt{provider} parameter is an example of cryptographically irrelevant parameter and will be identified as such.  

In addition, related crypto materials may be extracted from API's and parameters with in a similar fashion. Every type of related cryptographical material has its corresponding crypto semantic (e.g. a  private key, an initialization vector, a seed, salt, etc.), and is analyzed and associated based on the API, parameter value (if any) and parameter semantics.   
Finally, as explained above, the function that completes the cryptographical operation (e.g.  \texttt{Cipher.doFInal())} will be marked as a criterion for the extraction of cryptographic relevant slices.

Importantly, the association described for the \texttt{getInstance()} method above, does \textbf{not} include all the information needed to describe the complete cryptographic operation 
%i.e. the crypto asset which contains the \texttt{getInstance()} function
- demonstrating the need to identify related API calls. Specifically:
\begin{itemize}
    \item The actual operation, either "encrypt" or "decrypt" that is performed may be extracted from the information associated with the value of the first parameter in the \texttt{init(...)} function executed on the same \texttt{Cipher} instance. 
    \item The value of the initialization vector that was used (if any) for the "encrypt" operation may be extracted during the construction of the third parameter \texttt{ParameterSpec} of the \texttt{init(...)}, depending on functions used and parameter values being passed.
\end{itemize}

\section{Evaluation}
\label{sec:evaluation}

%In this section we evaluate \explorer abilities to discover cryptographic assets and showcase how the results can be used to construct a flow to detect and report cryptographic vulnerabilities.
In this section we evaluate \explorer and validate the following:
\begin{enumerate}
    \item \explorer is able to correctly and accurately identify the cryptography performed in real world applications.
    \item The analysis runtime is reasonable; meaning, the tool would be able to scan hunderds of applications in a matter hours. 
    \item The inventory of crypto assets can accurately convey the crypto \emph{semantics} and  support different use cases.
\end{enumerate}

This section is split into two parts. The first part, addressing the discovery of cryptography, is presented in section~\ref{sec:discovery_evaluation} and addresses points (1) and (2). The second, in section~\ref{sec:vulnerability_evaluation} addresses points (3). Specifically, we implement one of the use cases, vulnerability detection, on top of the inventory and evaluate and compare its effectiveness against state-of-the-art vulnerability detection tools.

\subsection{Discovery}
\label{sec:discovery_evaluation}

%In this section we evaluate \explorer abilities to discover cryptographic assets. 
%We start with presenting a set of open source projects which we use for evaluation, describe our methodology and results.

\begin{table*}[!ht]
    \centering
    \begin{tabular}{|l|c|c|c|c|c|}
        \hline
        \textbf{repository name} & \textbf{stars} & \textbf{forks} & \textbf{last modified} & \textbf{Java files} & \textbf{Java files with cryptography} \\
        \hline
        Mastercard/client-encryption-java   & 117 & 71 & ~1 month ago & 86 & 6 \\
        Peergos/Peergos                     & 2005 & 165 & ~1 month ago & 831 & 5 \\
        apache/httpcomponents-client        & 1467 & 973 & ~2 days ago & 775 & 5 \\
        andy-goryachev/PasswordSafe         & 18 & 0 & ~1 year ago & 689 & 7 \\
        hyperchain/javasdk                  & 67 & 37 & ~2 years ago & 395 & 11 \\
        wultra/powerauth-crypto             & 56 & 22 & ~2 months ago & 102 & 9 \\
        hyperledger/fabric-sdk-java         & 1119 & 708 & ~2 months ago & 300 & 6 \\
        \hline
    \end{tabular}
    \caption{Discovery reference open source projects. We provide the repository of each one of the reference projects, indication of its relevance (starts, forks and last modification date) as well as the number of overall java files containing cryptography.}
    \label{table:projects_metadata}
\end{table*}

\subsubsection{Experimental setup}

As no standard benchmark exists for crypto discovery, we construct a data set we believe is representative of real world application. Specifically, we identify seven java, open source, real world applications that exhibit different cryptographic usage patterns.
% These applications are used  which we use as a reference set to test \explorer. 
%(Note: \explorer does not aim to discover self-implemented cryptography.)
These open source projects, shown in table~\ref{table:projects_metadata} have been chosen for their popularity as well as inclusion of a variety of cryptographic operations. The table includes indication of each projects popularity, measured by both stars and forks, as well as information showcasing activity (last modified\footnote{Reviewed at the end of December 2024}), size (number of java files) and the number of java files containing cryptography.

Next, we semi-automatically analyze the data set and create a reference inventory. 
This process effectively labels our data set with expected crypto-assets. 
For simplicity, an asset in our manually labeled inventory will include only the most important information and cryptographic properties as follows:
\begin{itemize}     
    \item API location (file name, line number).
    \item The API itself (e.g., \texttt{javax.crypto.Cipher.doFinal}) 
    \item The cryptography algorithm being used (e.g., \texttt{AES}, \texttt{3DES}, \texttt{SHA-256}, etc.) 
    \item The cryptographic function (e.g. \texttt{keygen}, \texttt{digest}, \texttt{verify})
    % - one of the values defined in CBOM, including 
    % "generate",  (certificate generation, RNGs, ...)
    % "keygen",
    % "encrypt",
    % "decrypt",
    % "digest",
    % "tag",
    % "keyderive",
    % "sign",
    % "verify",
    % "encapsulate", and
    % "decapsulate". 
    \item The mode, if applicable (e.g., \texttt{CBC}, \texttt{GCM}, \texttt{ECB})
    \item The key size in bits, if applicable (e.g., 1024, 2048, 4096)
\end{itemize}

The process of labeling can be viewed as a 3 step process.
\begin{enumerate}
    \item \emph{Filtering.} In this step we filter all files in each of the projects to identify those that have cryptographic operations in them. This process can be accomplished, with relatively high accuracy, by "grep"-ing crypto relevant imports such as 
    \verb|java.security| or \verb|javax.crypto|

    \item \emph{LLM based labeling.} In this step we used LLM's, \emph{ChatGPT-4o}~\cite{openAI_website} and \emph{Mistral-Large-2407}~\cite{mistral_website}, to assist with generating crypto-assets. The LLM was provided with a file suspected of containing 
    cryptographic operations and was instructed to asses the file for the existence of 
    crypto and produce a Json summary. 
    
    %additional information generated to help improve accuracy was disregarded such as "container-%function" (the function/method containing the calls to the Cryptographic API )
    
    \item \emph{Manual labeling.} This step consists of the review of the LLM output, identify missing or errorneous crypto assets and validating that each crypto asset is complete and accurate. It is important to note that this step was found to be essential. The LLM labeling was performed without any tuning and as such produced results which we often used as a starting point. In many cases crypto-assets were partially identified, contained inaccuracies or even completely missing. 
\end{enumerate}

Overall, creating a reference inventory is a labourious task due to the LLM's low reliability with regard to identifying all crypto assets, cryptographic APIs and related properties. We posit that the LLM's performance could be enhanced through more rigorous prompt engineering and/or fine-tuning.
Moreover, manual labeling presented significant challenges as well. Labels were often refined during peer review processes and when our understanding of cryptography and APIs deepened. This peer reviewed, iterative labeling and refinement process increased our confidence in the quality of the labels.

Overall, at the end of the process our inventory contained 97 crypto-assets. Each asset represents cryptographic information contained within a single file; if, for example, a key is set outside the file and passed as a parameter, the key size property will not be present in the crypto-asset.
Table~\ref{table:projects_crypto_breakdown} shows a breakdown of the cryptographic assets identified across the different projects. 

\begin{table*}[!ht]
    \centering
    \begin{tabular}{|l|c|c|c|c|c|c|c|c|c|c|}
\hline
 \textbf{name} & \textbf{sign} & \textbf{digest} & \textbf {encrypt} & \textbf {verify} & \textbf{tag} & \textbf {keygen} & \textbf{key-} & \textbf {decrypt} & \textbf {en-/de-} & \textbf{total} \\
               &               &                 & \                 &                  &            
 &                  & \textbf{derive}    &                   & \textbf {capsulate} &     \\
 \hline
client-encryption-java  &   & 1 & 1 &   &   & 1 &   & 2 & 2 & 7 \\
Peergos                 &   & 4 &   &   & 3 &   &   &   &   & 7 \\
httpcomponents-client   &   & 9 &   &   &   &   &   &   &   & 9 \\
PasswordSafe            & 1 & 1 & 3 & 1 &   & 1 &   & 7 &   & 14 \\
javasdk                 & 7 & 5 & 4 & 3 &   & 7 & 2 & 6 &   & 34 \\
powerauth-crypto        & 2 & 3 & 1 & 1 & 3 & 4 & 1 & 1 &   & 16 \\
fabric-sdk-java         & 2 & 2 &   & 2 &   & 4 &   &   &   & 10 \\
\hline
\textbf{total}          & 12 & 25 & 9 & 7 & 6 & 17 & 3 & 16 & 2 & 97 \\
\hline
    \end{tabular}
    \caption{Discovery reference open source crypto asset breakdown. The table shows the number of different types of crypto assets - based on the crypto function - across each of our projects.}
    \label{table:projects_crypto_breakdown}
\end{table*}

%mention the advantages of our design and framework 
%improved separation of concerns, properties, (no need for pattern), ease of coding 
%ease of adding libraries and languages.

\subsubsection{Methodology}

We view the discovery task as a classification task and measure the true positive (TP) rate, false positive (FP) rate and false negative (FN) rate\footnote{We do not measure true negative as "everything" other than a crypto asset is considered a true negative}. Using these results we can determine precision and recall. 
%true positives (TP) are cases where the scanner was able to successfully detect a crypto asset that also appeared in the manual labeling.
%false positives (FP) are cases where the scanner incorrectly detected crypto assets that were not present in the manual labeling (because there was no crypto usage to be expressed or because they were different from what was found).  
%false negatives (FN) are cases where the scanner missed i.e. did not detect crypto assets that were present in the manual labeling.  

Next, we describe the criteria we use to match \explorer findings to our labels. 

%\textbf{Match  (true positives)}
\begin{itemize}
    \item \textbf{Exact Match}.
        An exact match occurs if all properties were found exactly as we expected in the label. specifically: 
        \begin{itemize}
            \item Identical code line numbers, API calls,  cryptographic function, and cryptography algorithms.
            \item Where applicable: Identical mode and/or key size~\footnote{Note: in cases where we could not label (manually resolve) a mode or keysize based on the single file, we assigned null. If in such case \explorer was able to find these properties in another file, we considered this a match. 
            This assumption was made to allow for reasonable labeling effort by avoiding analysing whole projects  to manually find callers for each such asset.}.
        \end{itemize}
    \item \textbf{Partial Match}.   
        A partial match occurs when the line and API call were correctly found but other crypto properties were missing or incorrect. 
        (A partial match, albeit with incomplete or inaccurate properties, would be beneficial for the developer as it would allow her to manually inspect and address any missing or inaccurate crypto properties)
\end{itemize}

It is important to note that our evaluation is qualitative. A keen observer will notice that a single (reference) label, crypto asset, assigned to a set of API calls in a file, may in fact represent multiple assets in different execution paths. Consider for example a function implementing decryption. In our reference inventory, a label will reference API calls in the function body. However, if a parameter to the function represents a key or a mode the label will not contain related properties such as key size or mode (as this information is not available in the file). In contrast, when considering an entire project - as does \explorer - multiple execution paths with calls to the decrypt function may be discovered - each with a specific key or mode. Thus, in our evaluation we try to match each label to at least a single asset, in a single execution path, \explorer generates. Multiple matches to a single label may exist - in such a case, we essentially "collapse" a set of matches to a single label.

\subsubsection{Accuracy}

The results of the evaluation are presented in  
table~\ref{table:crypto_performance_table}. The table shows, for each crypto function, the number of expected crypto assets, missed crypto assets (false negatives) as well as the number of matched and partially matched assets. In addition we provide the recall (based on full and partial matches) as well as false positives and precision. 

\begin{table*}[ht]
\centering
\begin{tabular}{|l|c|c|c|c|c|c|}
\hline
\textbf{crypto\_function} & \textbf{no. crypto assets} & \textbf{false negative} &         \textbf{match (partial)} & \textbf{recall (partial)} & \textbf{false positive} & \textbf{precision} \\
        \hline
        digest      & 25 & 1 & 24     & 96\%            &   & 100\% \\
        encrypt     & 9  &   & 7 (2)   & 78\%  (100\%)   &   & 100\% \\
        decrypt     & 16 &   & 13 (3)  & 81\%  (100\%)   &   & 100\% \\
        keygen      & 17 & 1 & 15 (1)  & 88\%  (94\%)    & 2 & 88\% \\
        verify      & 7  &   & 7      & 100\%           &   & 100\% \\
        sign        & 12 &   & 12     & 100\%           &   & 100\% \\
        encapsulate & 1  &   & 1      & 100\%           &   & 100\% \\
        decapsulate & 1  &   & 1      & 100\%           &   & 100\% \\
        keyderive   & 3  &   & 3      & 100\%            &   & 100\% \\
        tag         & 6  &   & 6      & 100\%           &   & 100\% \\
        \hline
        \textbf{total:}      & 97 & 2 & 89 (6) & 92\% (98\%)      & 2 & 97\% \\
\hline
\end{tabular}
\caption{Crypto asset discovery coverage. Each line represents crypto assets of a specific crypto function. We present the number of false negative (misses), correct and partial matches, recall (based on the matches), false positive and precision.}
\label{table:crypto_performance_table}
\end{table*}

Looking at the results, we can see that overall, \explorer is able to correctly identify 92\% of the assets and 98\% if we allow partial matches - while keeping a very low false positive rate.
A thorough analysis of the results reveals several reasons for misses or partial matches, chief among them are gaps in the data flow analysis. 
One example occurs when a value of an enumeration field is accessed and used as a parameter.
%or usage a variable that was initialized in a constructor and accessed at another function. 
Additionally, in our current implementation, we limit the static analysis to the application code. This means that dependent libraries are not analyzed and the analysis of the data flow in those paths is incomplete. This limitation is not inherent and can be removed, but will cost in increased run time.  
To summarize, overall, albeit the gaps that will be addressed in future versions, these results showcase the effectiveness of our tool and robustness of the design and implementation.

Lastly, \explorer results show a very high precision rate. The two false positives occur when \explorer assigns a crypto asset (key generation) to code effectively loading keys into memory. This is mainly a  matter of definition: during the labeling process we took a very narrow view and assign key generation only when a key bytes are randomly generated (e.g. using \texttt{keyGenerator.generateKey()}). Although this behavior can easily be modified by adjusting the knowledge base, in practice, we believe this information might be very useful to developers to understand which keys and from where those keys are loaded into the application.

\subsubsection{Runtime}

In this section we evaluate the performance of \explorer in terms of execution time.  
Our experimental setup includes a 64-bit Windows 11 running on a Intel Core I9 machine with 64GB of RAM. All of our experiments have been executed on this machine, with the goal of validating a reasonable runtime. Note that there are no stringent requirements on the speed of \explorer and we expect it to be executed offline most of the time.

\begin{table}[!h]
    \centering
    \begin{tabular}{|l|c|c|}
        \hline
        \textbf{name}                   & \textbf{lines} & \textbf{runtime (sec.)} \\
        \hline
        client-encryption-java & 8531   &  18 \\ % 00:00:18  \\
        Peergos                & 116297 &  40 \\ % 00:00:40  \\
        httpcomponents-client  & 127045 &  36 \\ % 00:00:36  \\
        PasswordSafe           & 109189 &  51 \\ % 00:00:51  \\
        javasdk                & 63565  &  52 \\ % 00:00:52  \\
        powerauth-crypto       & 14601  &  14 \\ % 00:00:14  \\
        fabric-sdk-java        & 54832  &  215 \\ % 00:03:35  \\
        \hline 
%        \textbf{Average} & & \\                        
    \end{tabular}
    \caption{Discovery execution time. The table shows the number of lines in each project and \explorer runtime measured in seconds.}
    \label{tab:execution_time}
\end{table}
% Average lines/sec:
% ( 127045 / 36 + 109189 / 51 + 54832  / 215 + 8531 / 18 + 116297 / 40 + 63565 / 52 +  14601 / 14 ) / 7 

Table~\ref{tab:execution_time} shows the run time of \explorer for each project.
The results show variations of scanning rates across different projects. This is due to the  filtering  \explorer employs to speed up the scanning. The filtering process removes non-crypto related files from the analysis. The filtering method essentially builds a call graph, identifies nodes where crypto API's are being called and prunes all other branches (the pruning is done at a file level granularity). 
Overall, the average speed observed was about 1650 lines per second. 
We believe this rate would be more than sufficient to perform scanning of hunderds or thousands of applications in large organizations within a reasonable time frame.

\subsection{Vulnerability detection}
\label{sec:vulnerability_evaluation}
In this section we show how we apply the inventory to identify crypto related vulnerabilities and provide state of the art accuracy when compared to specialized vulnerability detection tools.
We first provide a short review of our approach and then evaluate its effectivness.

\subsubsection{Approach}

Cryptographic vulnerabilities are often caused by lack of thorough understanding of the cryptographic assumptions and guarantees. Moreover, complex API's exacerbate the confusion and increase the likelihood of misuse. Common vulnerabilities include using insecure algorithms, insecure key generation or derivation as well as failure to provide sufficiently random nonces.

Identifying such vulnerabilities requires us to detect the cryptographic operation as well as all the related crypto material such as keys and nonces. This information is already captured by \explorer and encoded in every one of the crypto-assets in the inventory. As a result, a relatively simple set of rules could be applied on each crypto-asset to identify vulnerabilities. Note, that one of the benefits of our approach is the fact that these rules are encoded using only information provided in the inventory. Since the crypto-assets are language independent - the rules detecting cryptographic vulnerabilities are therefore language agnostic as well. In essence, the process of vulnerability detection is decoupled from the discovery process.

\begin{figure}[!ht]
    \centering
        \includegraphics[width=0.5\textwidth]{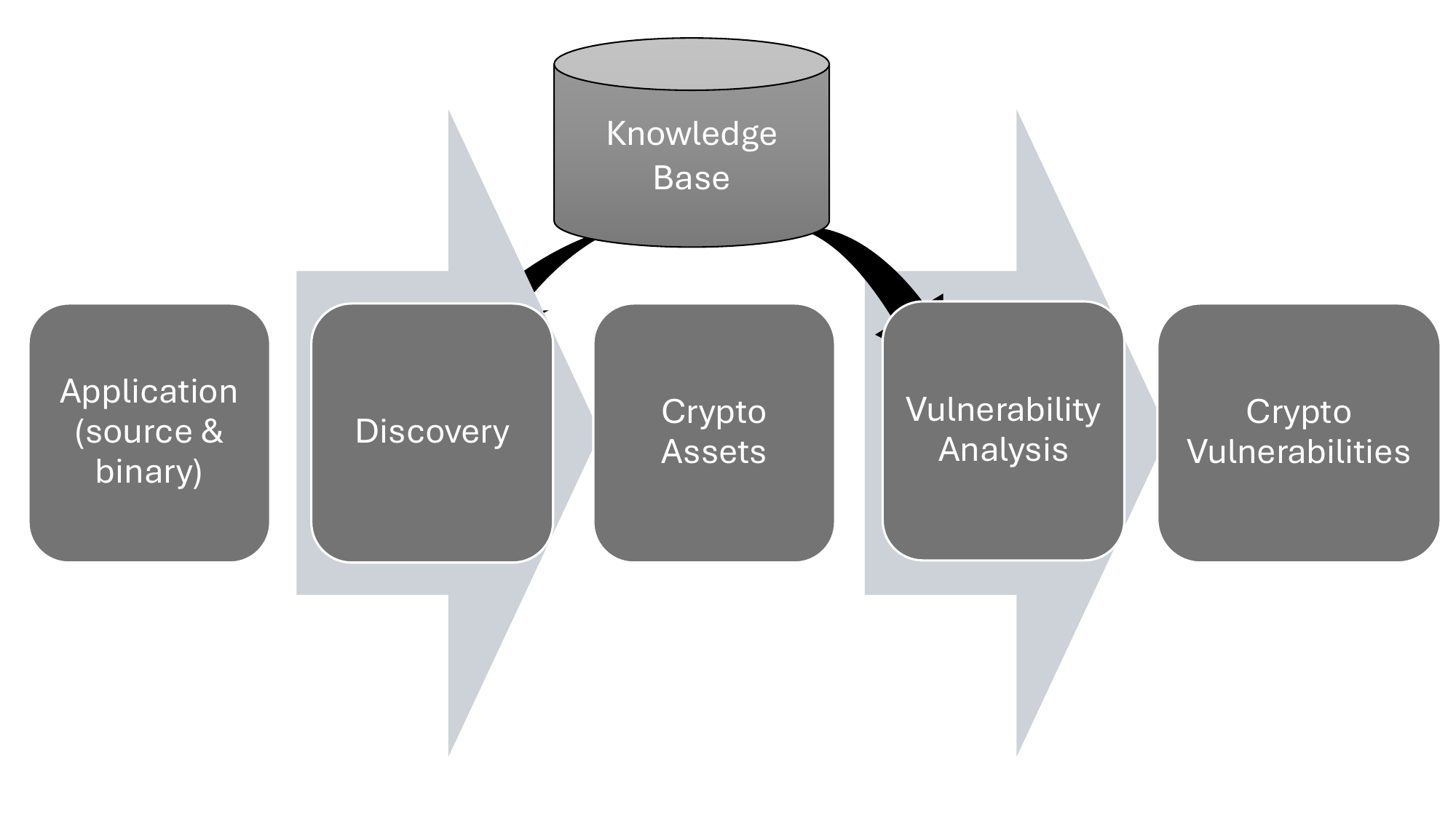}
    \caption{\label{fig:vulnerability-analysis} vulnerability identification flow
Crypto-asset discovery. The four stages of the static analysis pipeline: parsing the code, analyzing control and data flow, building slices based on crypto relevant criteria and finally constructing crypto assets for each slice.        
}
\end{figure}

The vulnerability analysis flow, shown in figure~\ref{fig:vulnerability-analysis}, uses the identified crypto assets and examines them for known vulnerabilities. 
The analysis is designed to be language-independent addressing only crypto vulnerabilities, so that a single analysis implementation 
will produce the same result over the usage of a broken cipher implemented in Java, Python, or other programming language. 
For example, it can assess the strength of a cipher based on its variant, its mode, the padding scheme, the key lengths, etc. Similarly, it can verify that the used algorithms are compliant with a given policy or if an algorithm is quantum-safe according to that policy e.g. by NIST ~\cite{nist_website} or the NSA ~\cite{nsa_website}  (i.e. CNSA 2.0 ~\cite{CNSA2}).
%The result of this analysis will contain the vulnerabilities that are discovered in each crypto asset.

Table~\ref{tab:explorer_vulnerabilities} lists the main vulnerabilities identified by \explorer. The table lists the related Mitre CWEs (Common Weakness Enumeration), the short descriptions as well as a the logic used to identify it based on the crypto asset\footnote{Note, CWE759 is appropriate in the context of password hashing and storing}. In addition, \explorer is able to identify more than a dozen "code smells" and unwanted patterns (their descriptions and functionality are beyond the scope of the paper).
%. For brevity these are not listed here.

\begin{table*}[!ht]
    \centering
    \begin{tabular}{|l|p{5cm}|p{10cm}|}
        \hline
        \textbf{CWE} & \textbf{Short Description} & \textbf{Analysis} \\
        \hline
%       CWE1204 may be valid, but should never happen in Java.
%       CWE1204 & Generation of Weak Initialization Vector (IV)                 & Identify crypto assets able to encrypt/decrypt, with modes that require IV and validate the related crypto material provided contains an IV and that is shorter than require size \\
        CWE259  & Use of hardcoded password                                    & Identify a password that is provided (as a parameter) to a crypto API function that has a (constant) value \\           
        CWE321  & Use of hardcoded Cryptographic Key                           & Identify a key, iv or salt that is provided (as a parameter) to a crypto API function that has a (constant) value \\
        CWE326  & Inadequate encryption strength                                & Check private key field size for small keys (in asymmetric algorithms) \\
        CWE327  & Use of a broken or risky cryptographic algorithm              & Identify weak and quantum unsafe algorithms in the asset variant, mode and key fields \\
        CWE328  & Use of weak hash                                              & Identify weak algorithms in variant field - for assets with a hash primitive  \\
%        CWE332  & Insufficient Entropy in PRNG                                  & Rule.. \\
        CWE335  & Incorrectly managed seed in Pseudo-Random Number Generator    & Identify a hardcoded (constant) seed that is provided (as a parameter) to a PRNG function \\
%        CWE336  & Same Seed in Pseudo-Random Number Generator (PRNG)            & Rule.. \\
%        CWE337  & Predictable Seed in Pseudo-Random Number Generator (PRNG)     & Rule.. \\
        CWE338  & Use of cryptographically weak Pseudo-Random Number Generator  & Validate the randomizer API call is one of the cryptographically strong PRNGs. \\
        CWE759  & Use of a one-way hash without a salt                          & Identify assets with primitives of type hash that do not contain related crypto material of type salt \\
 %       CWE760  & Use of a One-Way Hash with a Predictable Salt                 & Rule.. \\
        CWE780  & Use of RSA Algorithm without OAEP % (Bleichenbacher attack)  
                                            & Validate algorithm and padding scheme in the variant and padding fields \\
        CWE916  & Use of Password Hash With Insufficient Computational Effort   & Validate crypto API iteration parameter has a value larger than 1000 \\
        \hline         
        
    \end{tabular}
    \caption{The table shows the main vulnerabilities detected by \explorer and the logic  executed (on top of a crypto asset) to identify each one.}
    \label{tab:explorer_vulnerabilities}
\end{table*}

The vulnerabilities generated by \explorer are associated with a given crypto asset and contain all the relevant information. Specifically, they contains a unique id, reference to documentation from a trusted source (e.g Mitre CWEs or FIPS guidelines), reference to related crypto material, and list of the source code evidences.
% as defined by its unique id.
A vulnerability evidence is a reference to a crypto finding: % that contributes to the identification of the vulnerability. Crypto finding may be 
such as a crypto property, related crypto material, or an argument of a function call. 
The specific finding depends on the vulnerability type. %associated with the evidence.
Listing~\ref{listing:explorere_vulnerability} shows an example of the output produced by \explorer for a vulnerability caused by an unsafe algorithm which includes findings (evidences) of an API call - its location in the code and its argument value.
% \begin{listing}[t]
% \begin{minted}[
%     linenos=false,
%     frame=lines,           % Draw a frame 
%     fontsize=\small,      
%     breaklines=true,       % Allow breaking long lines
% ]{json}
% {
%   "vulnerabilityId": "2",
%   "classification": "cwe327",
%   "vulnerabilityScore": "Major",
%   "vulnerabilityDocumentationReference": "https://cwe.mitre.org/data/definitions/327.html",
%   "debugMessage": "Use of broken or risky cryptographic algorithm: DES",
%   "references": [
%     {
%       "type": "variant",
%       "value": "DES",
%       "context": {
%         "type": "FUNCTION_CALL",
%         "location": {
%           "fileName": ".../crypto/CipherUtil.java",
%           "line": 53,
%           "startColumn": 36,
%           "endColumn": 54
%         }
%       }
%     }
%   ]
% }
% \end{minted}
\begin{lstlisting}[language = Java, frame = tb, numbers=none,
captionpos=b,
caption = {A sample output from \explorer showcasing a vulnerability in the code. The result identifies the specific CWE, crypto properties, and location in the code.},
label = {listing:explorere_vulnerability}
]
{
  "vulnerabilityId": "2",
  "classification": "cwe327",
  "vulnerabilityScore": "Major",
  "vulnerabilityDocumentationReference": "https://cwe.mitre.org/data/definitions/327.html",
  "debugMessage": "Use of broken or risky cryptographic algorithm: DES",
  "references": [
    {
      "type": "variant",
      "value": "DES",
      "context": {
        "type": "FUNCTION_CALL",
        "location": {
          "fileName": ".../crypto/CipherUtil.java",
          "line": 53,
          "startColumn": 36,
          "endColumn": 54
        }
      }
    }
  ]
}
\end{lstlisting}
% \caption{A sample output from \explorer showcasing a vulnerability in the code. The result identifies the specific CWE, crypto properties, and location in the code.}
% \label{listing:explorere_vulnerability}
% \end{listing}

\subsubsection{Experimental setup}

%Vulnerabilities - this is more qualitative. Can't be apples to apples and can't compare to existing tools. 
%However, we should show some vulnerabilities are found and identified.

Our goal in this section is to compare the ability of \explorer to find vulnerabilities with that of existing tools. To achieve this, we use `CamBench`~\cite{schlichtig2022cambenchcryptographicapi}
%\footnote{`Cambench` can be found here: \url{https://github.com/CROSSINGTUD/CamBench/tree/main/CamBench_Real}}, 
- Cryptographic API Misuse Detection Tool Benchmark Suite for Java - as a third-party benchmark that enables us to evaluate \explorer and compare its capabilities to existing tools. 
CamBench, an active project aims to address the differences, strengths, and weaknesses of previously developed benchmarks for crypto misuse detectors e.g. CryptoAPI-Bench~\cite{CryptoAPI-Bench:8901573}, MuBench~\cite{MuBench/10.1145/2901739.2903506}, OWASP Benchmark~\cite{owaspBenchmark} etc., and combines both synthetic and real-world examples.
Cambench-Real~\cite{CambenchTool} covers 20 distinct GitHub projects and contains both secure and insecure API usages.
In particular, it contains 15 vulnerabilities spanning 10 repositories~\footnote{These repositories are real world applications that are mostly still in active use. Since some of the vulnerabilities may still persist in these projects, we are disguising specific identifying details, especially regarding the vulnerabilities, beyond what can be found in the original benchmark. 
}.

\begin{table*}[!ht]
        \begin{tabular}{|l|c|c|c|c|c|}
            \hline
             \textbf{ID} & \textbf{CogniCrypt} & \textbf{CryptoGuard} & \textbf{FindSecBugs} & \textbf{SonarQube}  & \textbf{\explorer}\\
            \hline
                 \textit{Insecure Key} &  & & & & \\
                 68-1  & \ding{51}  & \ding{55}   & \ding{55} & \ding{51} & \ding{51} \\
                 129 & \ding{51} & \ding{51} & \ding{55} & \ding{55}   & \ding{51}  \\
                 129-1 & \ding{55}  & \ding{51} & \ding{55} & \ding{55}   & \ding{51}    \\                
                 148 & \ding{51}  & \ding{51}  & \ding{55} & \ding{55}   & \ding{51}   \\
                 
                 \textit{Insecure Encryption Algorithm}   &    &   &   &  &    \\
                 68  & \ding{51}  & \ding{51}   & \ding{55}  & \ding{55}  & \ding{51} \\
                 146   & \ding{55}   & \ding{51}  & \ding{51}& \ding{51}& \ding{51} \\
                 \textit{Insecure Initialization Vector} &   &    &  &  &   \\
                 71-1   & \ding{51} & \ding{55}  & \ding{55}  & \ding{51}   & \ding{51} \\
                 \textit{Insecure Message Digest}   &   &  &   & &  \\
                 34-2  & \ding{51}  & \ding{51} & \ding{51} & \ding{55} & \ding{51}       \\  
                 \hline
                 \textbf{Total subset (8)}  & 6  & 6  & 2    & 3 & 8 \\
                 \hline
                 \textit{Insecure Mac Algorithm}  &  &  &  & &  \\
                 110  & \ding{51}  & \ding{55} & \ding{55}  & \ding{55}   & \ding{51} \\
                 \textit{Insecure Password-based Encryption} &   &  &   &  &  \\
                 72  & \ding{51}  & \ding{55}  & \ding{55} & \ding{55} & \ding{51} \\
                 \textit{Insecure Signature Algorithm} &  &   &    &   &  \\
                 45 & \ding{51}    & \ding{55}   & \ding{55} & \ding{55} & \ding{51}  \\
                 \textit{Usage of String for Sensitive} Information & &  & & &  \\
                 33  & \ding{51} & \ding{55} & \ding{55} & \ding{55}  & \ding{55}  \\
                 45-1   & \ding{51}  & \ding{55}  & \ding{55} & \ding{55} & \ding{55}  \\
                 99  & \ding{51}  & \ding{55} & \ding{55}  & \ding{55} & \ding{55}  \\
                 151  & \ding{51}  & \ding{55} & \ding{55}  & \ding{55} & \ding{55} \\
                 \hline
                 \textbf{Total all (15)}  & 13  & 6   & 2   & 3  & 11 \\
            \hline
        \end{tabular}
         \caption{This table extends the table in~\cite{CambenchTool} by adding a column for \explorer on the right. The table shows which vulnerabilities were discovered by each one of the tools.}
         \label{table:cambench_results}
\end{table*} 

The benchmark was used to compare CogniCrypt~\cite{CogniCrypt},
% \footnote{\url{https://www.cognicrypt.org}}, 
CryptoGuard~\cite{CryptoGuard:10.1145/3319535.3345659}, 
SpotBugs plugin~\cite{SpotBugs}, and 
SonarQube~\cite{SonarQube}. We extend the analysis by adding \explorer into the comparison.
Table~\ref{table:cambench_results} shows the extended table as presented in~\cite{CambenchTool} with an additional column for \explorer (right most column). Note, the ID column, uniquely identifies an API call within a repository.
%, can be disregarded for the purposes of this article.

\subsubsection{Accuracy}

The authors of Cambench split the vulnerabilities into two subsets. In the first subset, our tool outperformed the other tools by finding eight out of eight vulnerabilities, giving \explorer the only perfect score. 

In the second subset, only CogniCrypt %, which was developed by the same group,
%is CamBench's flagship product, 
was able to find all vulnerabilities. 
%They separated out that subset in order to sidestep conflict of interest issues.
In this subset \explorer was able to identify 3 vulnerabilities while missing 4 instance, all of type "\textit{Usage of string for sensitive information}". The vulnerability manifests when sensitive information resides in memory more (time) than required. In java, this may occur if sensitive information is stored in a String (instead of char[]). Since java Strings are immutable and can't be overwritten - the sensitive information resides in memory until it is reclaimed by the garbage collector. We consider this vulnerability a security vulnerability rather than a cryptographic one and as such is not within our threat detection scope. Note, that a cryptographic asset was identified in all these cases (without a vulnerability).
% I am not sure we want to say we can address this - as this information is not present in a crypto asset.
%Overall, \explorer scored 10/15 and was bested only by CogniCrypt.
To summarize, \explorer was able to correctly identify all crypto related vulnerabilities. 

%TODO: explain that given \explorer`s abstract nature, this is an impressive feat  

% What is the point of the below?
% - I attempted to show that Explorer has potential beyond the scope of CamBench vulnerabilities..
%It is also worth pointing out that CamBench focuses on the most common vulnerabilities and security breaches while \explorer finds a large variety of vulnerabilities and warnings from "Use of AES Algorithm with key size less than 256 bits" to "Use of non-quantum-resistant public-key algorithms." 

%TODO: Add more information about the difference between our tools, defining our general solution vs the other tools being more specific 

\section{Related work}
\label{sec:relatedArt}

In this section, we focus on related art that is applicable to discovery of cryptography as well as vulnerability and misuse identification (in the context of software applications). 
Works related to network monitoring, cipher suites, network protocols, certificates, etc.~\cite{Qualys, TestSSL,SSHScanServer, SSHScanCipher, IkeScan} 
do not relate directly to our work.
%and are considered out of scope.

It is important to note the distinctions between discovery and vulnerability/misuse detection. The majority of works do not make such a differentiation as creating an inventory was not a goal by itself. Indeed, the process of vulnerability or misuse detection implicitly detected crypto elements. However, consider the following differences between discovery and vulnerability/misuse detection tools.
\begin{itemize}
    \item In cases where there are no vulnerabilities in the code, vulnerability/misuse detectors will result in an empty set. Essentially no crypto will be identified by the vulnerability identification tools. 

    \item Whenever a vulnerability is identified all the information that pertains to the vulnerability is provided. However, there is no requirement or need to provide complete information about the cryptographic operations. For example, consider a case where a vulnerability related to a hard coded IV is discovered and identified. Such a result can provide very limited information about the cryptographic operation as a whole, including for example the algorithm, key, mode, function etc.
\end{itemize}

%Clearly differentiate between discovery and vulnerability identification. The reader should understand our goal is not comparable to previous work

\subsection{Crypto discovery}

Most of the work related to crypto discovery and cryptographic inventory has been performed by the industry. 
InfoSecGlobal ‘AgileScan’~\cite{AgileSec} and SandboxAQ AQtive Guard~\cite{AQtiveGuard} provide
methods to discover cryptography in applications. The former is able to identify cryptographic objects such as certificates, keys % SSL Certificates
and crypto libraries that are used. The latter monitors applications during run time for known API calls to cryptographic libraries. 
Both, to the best of our knowledge, are not able to collect enough semantics to construct a complete crypto asset. 

The Wind River open source crypto detector~\cite{WindRiver} scans code looking for keywords such as “RSA”, known usage patterns such as API calls, data type declarations or \texttt{\#include} statements to construct a list of cryptographic occurrences. CodeQL~\cite{CodeQL} is a static analysis engine. It allows users to query code represented in a relational database. CodeQL collects information, for each source code, such as the abstract syntax tree, bindings and type information as well as data and control flow. 
%CodeQL support most common languages however queries are language specific. 
The work described in~\cite{CodeQLCbom} builds on top of CodeQL to discover and generate an inventory (in the form of a CBOM) for open-source git hub projects. The work in~\cite{CodeQLGithub} lays out some of the processes used to generate the CBOM. The process builds on CodeQL results and in particular the text snippets found. It then uses regular expressions to identify algorithm names, key sizes, block cipher mode etc., which are encoded in a structure. Next, a CBOM component is constructed using all the properties that were matched or could have been derived. 
%(e.g. by encoded mapping primitives ‘rsa: pke’). 
Finally, there is a process for merging overlapping components when necessary. 

The goal and approach of the latter work has a few similarities to our own, however, there are a few points we believe our contributions address. First, the work describes in~\cite{CodeQLGithub} uses regular expressions which contain language-specific statements (in fact, library-specific). Instead, \explorer separates the algorithm from the meta-data. (The meta-data resides in the knowledge base which may be augmented without changing the code.) Secondly, \explorer leverages control and data dependencies to detect and relate cryptographic API's. Without it, we believe inaccuracies may fall if code is complex and interleaves multiple types of cryptography that can not be correctly separated using regular expressions. Lastly, it is not clear what contexts are accounted for in~\cite{CodeQLGithub} - which may impede the ability to relate cryptographic components across the code - and provide a complete description of the cryptographic operation. Such a case may occur, for example, when a key is generated in one method/file and used during decryption in another method/file. 

An additional effort, closely related to our own, is a plugin to sonar cube~\cite{sonar-cryptography}. 
This plugin can identify cryptographic assets within source code and create a CBOM. It operates on a rule-based system, where specific rules are established to recognize cryptographic APIs and their properties. Additionally, related cryptographic APIs can be detected by defining extra dependent rules, which is similar to outlining various patterns of sequences for cryptographic APIs. Note, a unique set of detection rules is required for each programming language as well as additional set of dependent rules for  each crypto library.
These are the specific limitations that \explorer addresses by (1) utilizing data and control flow analysis and (2) storing all cryptography-related information separately in the knowledge base (KB).

\subsection{Crypto vulnerability detection}

In this section we review related work addressing crypto vulnerability (misuse) detection. These examples are provided as background for section~\ref{sec:vulnerability_evaluation}. A more thorough review can be found here~\cite{TaintCrypt}. 
 
Previous research shows that crypto is often used in an insecure 
way~\cite{EvaluationInCryptoAndroid/10.4108/eai.3-12-2015.2262471, CryptoLint/10.1145/2508859.2516693, WhyCryptoSWFail/10.1145/2637166.2637237,WhyJavaDevStruggle}. One such problem is the choice of an insecure parameter, like an insecure block mode, for crypto primitives like encryption. Many tools exist to identify these misuses such as CryptoREX~\cite{CryptoREX.242030}, CryptoLint ~\cite{CryptoLint/10.1145/2508859.2516693}, CrySL~\cite{CrySL.validatingcorrectusage}, CogniCrypt~\cite{CogniCrypt}, and Cryptoguard~\cite{CryptoGuard:10.1145/3319535.3345659}. 

The authors of LICMA~\cite{LICMA} present 6 rules (5 in Python) aimed at finding vulnerabilities. These rules require the identification of the API and the parameter(s) to the API. To this aim, static analysis, specifically backward slicing is used starting from the relevant API parameters and going backward to determine if the value of a parameter was hardcoded, assigned locally or globally, and if possible provide the value. To the best of our knowledge the authors are able to trace parameters across function boundaries - but due to the underlying parser - are not able to trace parameters across file boundaries. 

Similarly, TaintCrypto~\cite{TaintCrypt} makes use of data flow and taint flow analysis to identify vulnerabilities using rules that can be modeled as DFAs (Deterministic Finite Automata). It is important to note that TaintCrypt works on C/C++ only and requires a configuration identifying sources, sinks and filters (which may not be trivial to identify). Moreover, the majority of the vulnerabilities addressed are aimed at the implementation of the crypto itself, while only a few are aimed at the API usages.

CryptoLint~\cite{CryptoLint/10.1145/2508859.2516693} is a light-weight static analysis tool designed to identify common cryptographic API misuse in Android applications that may lead to vulnerabilities such as weak encryption modes, insecure randomness, or improper key derivation. In particular, the tool scans for vulnerabilities defined in six rules. cryptoLint~\cite{CryptoLint/10.1145/2508859.2516693} also suggests remediation measures to help address the widespread issues identified.

CrySL~\cite{CrySL.validatingcorrectusage}, a definition language that enables cryptography experts to specify the secure usage of the cryptographic libraries that they provide. CrySL combines the generic concepts of method-call sequences and data-flow constraints with domain-specific constraints related to cryptographic algorithms and their parameters, return values, and state transitions, ensuring that developers adhere to secure coding practices. CogniCrypt~\cite{CogniCrypt} is a compiler that translates the CrySL~\cite{CrySL.validatingcorrectusage} rule set into a context and flow-sensitive static analysis program for Java or Android apps able to check (and also provides actionable guidance) for violations of the CrySL encoded rules.

The work described in~\cite{CDRep} addresses both the detection of crypto misuse as well as remediation. CDRep addresses 7 vulnerabilities, in machine language (java byte code).
%using 7 rules. 
The process to identify a vulnerability involves a starting point (e.g. the getInstance instruction) as well as a process detecting data dependencies (e.g. of the parameters of the instruction). In cases where the parameter can be resolved to a value (e.g. constant) a vulnerability can be identified. 

A different approach, using instrumentation of java JCA classes to log parameters of relevant Crypto APIs calls is presented in~\cite{CryLogger}. Once the execution of the program is complete, and the parameters are logged, a checker reviews the logs and enforces %predefined 
rules. The vulnerabilities covered by this approach can be extended to include rules that track behaviour across multiple executions of the program, such as reuse of key/iv. The work presented in~\cite{iCryptoTracer} also employs static analysis to identify and monitor Crypto API calls as well as file and network I/O API calls during run time. They are collecting the calls, parameter values and return values to detect, using a set of rules, whether cryptography was improperly employed on data sent to the network or stored on disk.

CRYScanner~\cite{CryScanner} combines both static and dynamic analysis perspectives. It uses an online logger, CRYLogger~\cite{CryLogger} and an offline checker, CrySL and CogniCrypt~\cite{CogniCrypt}. The tool runs in two stages, the runtime stage to collect logs and the offline stage to analyze these logs. It supports several popular Java cryptographic libraries and leverages domain-specific knowledge to pinpoint misuse patterns effectively. %An important goal of CRYScanner~\cite{CryScanner} is to reduce common cryptographic errors by enabling developers to follow security best practices during the development process. 

% Removed due to space. We can return but it is not providing much new information
%%%%%% 
%Another approach is present in~\cite{SMV-Hunter}. This work aims to identify SSL man-in-the-middle vulnerabilities caused by developers who, for example, avoid validation or simply allow all certificate and/or hostnames. The authors use both static and dynamic analyses, first to identify whether specific interfaces are overrided and known patterns (such as 'no-op') are implemented, and second to execute the application and monitor its behaviour during an attack. 

To summarize, we can identify two characteristics most works exhibits: first, vulnerability identification rules are language specific and are therefore not applicable across different languages. Second, most vulnerabilities are detected by identifying parameter values of select crypto API's (during run time or using static analysis data tracing) without an explicit aim - nor success - of understanding the complete cryptographic operation. CRYLogger~\cite{CryLogger} and CRYScanner~\cite{CryScanner} are in fact, dividing the vulnerability/misuse detection into separate stages, however, they also are not providing a generic expression of the used cryptography.

\section{Conclusion and Future work}
\label{sec:summary}

In this paper we present \explorer - an industrial grade tool for discovering cryptography in applications. Organizations will be able to deploy \explorer on large code bases and will be provided with a fast, accurate, high level and comprehensive view of the cryptographic operations performed in the code. Moreover, \explorer covers a large set of crypto related vulnerabilities and detects with high accuracy more cryptograhically related vulnerabilities than any other tool we are awere of. 
%on top of crypto-assets and showcase how we can decauple vulnerability identification from the discovery process and in particular from a specific programming language.

Next, we plan to progress in two main directions. First, we plan to make use of \explorer to set the stage for remediation suggestions. By building on top of the results generated by \explorer we would be able to pinpoint locations in the code where vulnerable or misused cryptography exists and utilize this information to inform the process of remediation whether it is LLM- or rule-based. Second, we are looking to expand the support to additional common programming languages, such as Python, Go, C etc..

\section*{Acknowledgments}
We would like to acknowledge the help given by the authors of Cambench~\cite{CambenchTool} and in particular Dr. Anna-Katharina Wickert for providing us access to the git repositories that were used to produce the evaluation metrics.

\bibliographystyle{plain}
%\bibliography{\jobname}
\bibliography{main}

%%%%%%%%%%%%%%%%%%%%%%%%%%%%%%%%%%%%%%%%%%%%%%%%%%%%%%%%%%%%%%%%%%%%%%%%%%%%%%%%
%\end{document}
%%%%%%%%%%%%%%%%%%%%%%%%%%%%%%%%%%%%%%%%%%%%%%%%%%%%%%%%%%%%%%%%%%%%%%%%%%%%%%%%

% original    
%\bibliographystyle{alpha}
%\bibliography{sample}

\end{document}